\documentclass[12pt]{iopart}
\usepackage{textcomp}
\usepackage{hyperref}
\usepackage[dvips]{graphicx}
\usepackage{delarray}
\usepackage{amssymb}
\usepackage{color}
\usepackage{url}

\newcommand{\Bv}{\mathbf{B}}
\newcommand{\bv}{\mathbf{b}}

\newcommand{\cd}{\cdot}

\newcommand{\fz}{f_0}
\newcommand{\fm}{f_m}
\newcommand{\fj}{f_j}
\newcommand{\fzm}{f_{m0}}
\newcommand{\fzj}{f_{j0}}

\newcommand{\gs}{{\sc gs2}}

\newcommand{\het}{$^{3}\rm He$}

\newcommand{\na}{\nabla}

\newcommand{\p}{\partial}

\newcommand{\teff}{T_{\rm eff}}

\newcommand{\vv}{\mathbf{v}}

\newcommand{\vpe}{v_\perp}
\newcommand{\vpa}{v_\|}
\newcommand{\vve}{\mathbf{v}_{\mathbf{E}}}

\begin{document}

\title[]{Turbulent transport of MeV range cyclotron heated minorities
  as compared to alpha particles}

\author{Istv\'{a}n~Pusztai$^{1}$, George~J.~Wilkie$^{1,2}$, 
Yevgen~O.~Kazakov$^{3}$ and T\"{u}nde F\"{u}l\"{o}p$^{1}$}

\address{$^1$ Department of Physics, Chalmers University of
  Technology, SE-41296 G\"oteborg, Sweden}

\address{$^2$ Institute for Research in Electronics and Applied
  Physics, University of Maryland, College Park, MD 20742, USA}

\address{$^3$ Laboratory for Plasma Physics, LPP-ERM/KMS, 1000 Brussels,
  Belgium}

\ead{pusztai@chalmers.se}

\begin{abstract}

We study the turbulent transport of an ion cyclotron resonance heated
(ICRH), MeV range minority ion species in tokamak plasmas. Such highly
energetic minorities, which can be produced in the three ion minority
heating scheme [Ye. O. Kazakov et al. (2015) \emph{Nucl. Fusion} {\bf
    55}, 032001], have been proposed to be used to experimentally
study the confinement properties of fast ions without the generation
of fusion alphas. We compare the turbulent transport properties of
ICRH ions with that of fusion born alpha particles. Our theoretical
predictions indicate that care must be taken when conclusions are
drawn from experimental results: While the effect of turbulence on
these particles is similar in terms of transport coefficients,
differences in their distribution functions -- ultimately their
generation processes -- make the resulting turbulent fluxes different.

\end{abstract}

\maketitle


\section{Introduction}
\label{intro}

The success of the magnetic confinement approach to fusion strongly
relies on the confinement and transport of the alpha particles over
the whole energy range from $3.5\,\rm MeV$ to the thermalized
ash. These particles represent the source of heat for the
self-sustained fusion reaction, they can damage the plasma facing
components if they get lost before slowing down, and they can dilute
the plasma if they would not leave the core after depositing their
energy.  Accordingly, a significant theoretical and experimental
effort has been dedicated to the study of alpha particle physics
\cite{iter1, iter2}. 

The ITER experiment is expected to demonstrate dominant alpha heating
for the first time. In order to gain confidence in our
  capabilities to predict the behavior of alphas, various experiments
  and proposals considered mimicking alpha particles using energetic
  ion species from neutral beams and/or radio frequency (RF) heating
  (see for instance \cite{mantsinen02,pizzuto10}). Similarity scaling
  arguments can provide a guidance for such experiments to be relevant
  for alpha physics in reactor scale devices
  \cite{pizzuto10,lackner98, romanelli04,calabro09}.  Previous works
  have mostly been concerned with creating ITER-relevant situations in
  devices of smaller dimensions. Since ITER will operate in a
  non-activated mode for several years before starting to generate
  alphas in deuterium-tritium (DT) plasmas, it is instructive to
  consider possibilities to mimic alpha particles in scenarios which
  can be applied on ITER itself.  For such experiments the generation
  of energetic ions in the MeV range would be desirable, which is not
  trivial to achieve using conventional heating techniques. In this
  paper we theoretically demonstrate that certain aspects of the
  \emph{turbulent} transport of alpha particles may be studied already
  in the non-activated phase through generating very energetic trace
  minority ions by a novel ion cyclotron resonance heating (ICRH)
  scheme, while care must be taken with the interpretation of any
  experimental results because of the different nature of these
  species.

In this paper we only consider tokamaks, where, due to the almost
perfect toroidal symmetry the vast majority of collisionless orbits
are confined. Imperfections in toroidal symmetry can lead to
  direct orbit losses; it is a well understood process that can be
  effectively mitigated using ferritic steel inserts \cite{tobita03}.
Fast particle driven instabilities -- in particular various
Alfv\'{e}nic Eigenmodes (AE) -- have been playing a major role in
energetic particle losses in current tokamak experiments, and
accordingly, they have received considerable attention (see
recent reviews in Refs.~\cite{gorelenkov14,chen16}). The importance of
AEs in ITER is still an open question; a recent study
\cite{pinches2015} concludes that AE-induced transport is not expected
to play a major role below mid-radius in a typical ITER scenario (due
to the Landau damping being stronger than the fast particle drive).
In any case, in a reactor the energetic alpha particle losses need to
be kept within some tolerable level and for the majority of the alpha
particles the collisional slowing down should happen on a flux
surface; in this paper we assume this to be the case.

The effect of turbulence on fast ions is suppressed by finite Larmor
radius (FLR) effects at very high energies. However, as pointed out in
e.g.~Ref.~\cite{wilkie}, across some suprathermal energy range
turbulence can play a major role in radially transporting alpha
particles; in particular, the radial turbulent transport timescale can
be much shorter than the slowing-down timescale. Thereby the energy
distribution of alphas may be modified from the usually assumed
slowing down distribution \cite{wilkie16,wilkiethesis,sigmar93}.

The energetic species that is used to mimic alphas should optimally
reach temperatures well above the critical energy for electron drag to
dominate the collisional slowing down \cite{pizzuto10}. There have
been experimental studies generating high-energy ions to simulate
fusion-born alpha particles. For instance, a neutral beam injection
hot ion population of ${}^4$He was further energized from the
$100\,\rm keV$ to the $\rm MeV$ range using third harmonic ICRH
\cite{mantsinen02}.  Recently, another possibility to generate
energetic ions with ICRH has been proposed theoretically \cite{yevgen}
and observed indirectly in experiments \cite{wukitchAPS}. A distinct
feature of this three-ion minority (TIM) heating scenario is the high
efficiency of the power absorption at a very low concentration of the
resonant ions. As a result, the minority ions can be accelerated to
higher energies than in commonly used heating scenarios. Here we
consider one of the possible TIM scenarios in ITER \cite{yevgen}, and
compare the transport properties of the heated minority to that of the
alpha particle transport in a similar DT discharge. The strong
non-Maxwellian feature of the energetic trace species is taken into
account in our gyrokinetic analysis. We find that although turbulence
advects these trace species in a similar way -- apart from minor
differences due to ionic composition effects -- their transport
properties can be very different because of differences in their
distribution function. Alphas tend to transport outwards in steady
state, while turbulence acts to accumulate the heated minorities in
the region where the power absorption is the strongest.

Besides its relevance in fusion, the transport of high energy (and
non-Maxwellian) minority species is also of interest in space- and
astrophysics context. For instance, in the solar wind plasma the
effective temperature of various ions have been observed to increase
linearly or stronger with atomic mass \cite{schmidt1980}, which may be
a result of turbulent heating of heavy ions \cite{barnesheating} in
this highly collisionless environment. The methods and some of the
results, presented here are also relevant in such circumstances.

The remainder of this paper is organized as follows. In
Sec.~\ref{heatedspecies} we discuss the heating scenario considered,
and explain the modeling of the distribution function of the heated
species. Sec.~\ref{turb} starts with general considerations on how the
turbulent transport of a trace species is treated, which is followed
by the setup of the gyrokinetic simulations in
Sec.~\ref{gkmodel}. Then we present some general observations for the
transport of a hot species in Sec.~\ref{fluxgrad}, before getting to
the main results: comparing the transport of heated minorities with
that of alpha particles in Sec.\ref{transportresults}. Finally we
conclude in Sec.~\ref{conclusions}.

\section{Scenario and properties of the heated species}
\label{heatedspecies}

\begin{figure}
\centering
\includegraphics[width=0.5\textwidth]{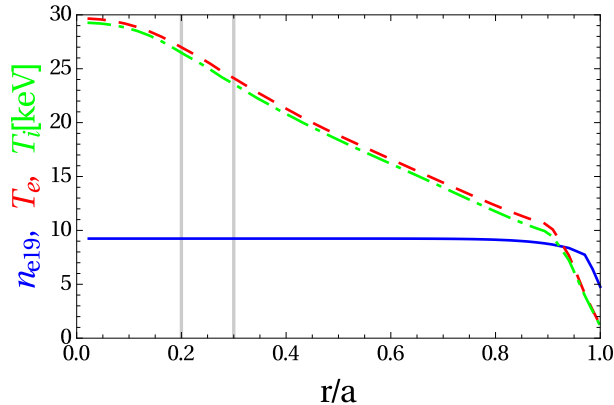}
\caption{Density ($n_e$ in $10^{19}\,\rm m^{-3}$, solid curve) and
  temperature profiles ($T_e$ and $T_i$ in $\rm keV$, dashed and
  dash-dotted curves, respectively) in the projected ITER discharge
  20020100. The radial region studied is indicated by the vertical
  bars.}
\label{profilefig}
\end{figure}

We are interested in to what extent the transport of alpha particles
in DT plasmas can be mimicked by that of heated minority ions in
plasmas with different ion composition, but similar profiles. In
particular, we consider a non-activating ion composition: hydrogen
($69.8\%$ of electron density) and helium-4 ($15\%$), with helium-3
($0.1\%$) minority as the heated species. The choice of the profiles
is based on projected profiles in an ITER scenario simulation
\cite{roach}. The profile data was accessed from the International
Multi-tokamak Confinement Profile Database.  We choose a high
$\beta_N$ hybrid \cite{gormezano07} discharge (20020100), the
temperature and electron density profiles of which are shown in
Fig.~\ref{profilefig}. The profiles represent the self-consistent
solution of an interpretative transport analysis using the PTRANSP
code \cite{budny}, as detailed in the Appendix of
Ref.~\cite{roach}. We kept the electron density, and assumed the
concentration of the different ion species to be constant.
Furthermore, some separation of the bulk ions might occur due to mass
and charge effects on their turbulent transport
\cite{estradamila05,isotope}, and also modifications of the magnetic
equilibrium when moving between the different ion composition cases,
but these changes are not expected to qualitatively affect our
results.

For our baseline study we will use the same temperature profiles and
electron density profile as in the projected DT discharge. In the
non-burning phase of ITER it is unlikely that such temperatures could
be reached, but we will show in Sec.~\ref{transportresults} that the
energy dependence of the radial transport remains similar apart from
scaling with the bulk temperature. In the simulations the hot minority
is assumed to have a radially constant concentration, unless stated
otherwise, and have distribution functions unaffected by the energy
dependence of the radial transport.

The location of the ICRH resonance affects the achievable effective
minority temperature. More central location leads to hotter minorities
due to the absorption being localized to a smaller volume. Therefore
generating energetic ions with central heating requires less ICRH
power and may be advantageous to use in an experiment. However, here
we choose a somewhat off-axis resonance location to assist our
gyrokinetic study. Gyrokinetic simulations often fail to predict a
finite turbulent transport close to the magnetic axis (see for
instance \cite{holland08}), while some level of turbulent transport
would be needed to predict temperature profiles consistent with the
experiment. Without going into the discussion of possible reasons for
this observation of turbulence simulations, we simply note that it is
advantageous to choose a resonance location tangential to a flux
surface where microinstabilities with positive growth rates were
present in order to get a finite level of turbulent fluxes in the
simulation. Accordingly, we choose the cyclotron resonance of \het{}
ions to be tangential to the $\rho=0.3$ flux surface from the outboard
side ($\rho=r/a$ is the normalized radius, $r$ is the radial
coordinate defined as the half width of the flux surface at the
elevation of its centroid, and $r=a$ at the last closed flux
surface). For this resonance location and an on-axis toroidal field of
$B_T=5.3\,\rm T$ the RF source should operate at the frequency $f_{\rm
  ICRH}=50\,\rm MHz$. A single toroidal wave number $n_{\rm tor}=27$
representative for $[0,0,\pi,\pi]$ phasing of the ITER ICRH antenna
\cite{budny12,dumont09} is used in the simulations.

\begin{figure}
\centering
\includegraphics[width=0.49\textwidth]{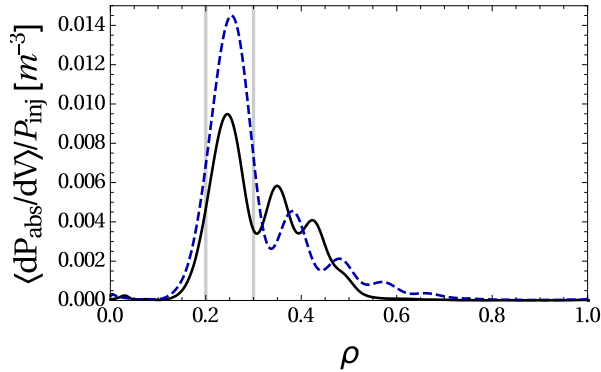}
\caption{Flux surface averaged RF power density per injected power
  $\rm [MW/(m^3 MW_{ inj})]$, absorbed by $^{3}\rm He$ ions; from
  TORIC simulations. Solid curve: at baseline profiles (shown in
  Fig.~\ref{profilefig}); dashed curve: at reduced temperature and
  density (profiles scaled from those in Fig.~\ref{profilefig},
  corresponding to $T_i(0)=4\,\rm keV$ and $n_e(0)=6\cdot 10^{19}\,\rm
  m^{-3}$). Note the different scales.}
\label{deposition}
\end{figure}

The flux surface average absorbed power density per unit injected
power for \het{} ions computed with the TORIC code
\cite{brambilla,bilato} is represented by the solid curve in
Fig.~\ref{deposition}. The simulations were made for the baseline
plasma parameter profiles, assuming Maxwellian distributions for all
particle species, and an effective temperature of $1\,\rm MeV$ for
\het. The heating modeling does not account for finite orbit width
effects and does not evolve the distribution functions and the power
deposition self-consistently. While a detailed ICRH modeling is
outside the scope of our conceptual study, we note that such
effects may lead to radial power deposition profiles which are
considerably smoother, corresponding to possibly less sharp variations
in the effective temperature of the heated minority. The
  importance of the relative magnitude of the density and temperature
  gradients of the hot species is discussed in
  Sec.~\ref{transportresults}.

A similar calculation for reduced temperatures and electron density is
represented by the dashed curve in Figure~\ref{deposition}; the
profiles are re-scaled so that the central ion temperature and
electron density are $T_i(0)=4\,\rm keV$ and $n_e(0)=6\cdot
10^{19}\,\rm m^{-3}$, respectively. In this case, when the bulk is
colder, electron absorption is not efficient and almost $100\%$ of the
coupled power is absorbed by the \het{} ions.  For the baseline case,
$23\%$ of the launched power is absorbed by electrons with a radially
broad absorption profile at the low field side. To reach similar
minority-to-bulk temperature ratios in the studied radial region, we
assume different injected ICRH powers: $P_{\rm inj}=15\,\rm MW$ in the
baseline case and $P_{\rm inj}=10\,\rm MW$ for the re-scaled profiles.

The ICRH produces anisotropic minority distributions with a
perpendicular-to-parallel temperature ratio larger than
unity. However, with our gyrokinetic tool we can only model an
isotropic species.  This simplification is expected to have some
effect on the energy dependent particle fluxes. It has been shown
\cite{hauff08,pueschel12} that, in electrostatic turbulence, at very high
energies $E$ the radial diffusivity of particles scale as $E^{-3/2}$
for most of the particles except those with $|\vpa|/v$ very close to
$1$. (We introduced the particle speed $v=|\vv|$ and the parallel
velocity $\vpa=\vv\cd\bv$, where $\vv$ is the particle velocity, and
$\bv=\Bv/B$, with $B=|\Bv|$ and $\Bv$ the unperturbed magnetic field.)
Therefore, at least in the high energy limit, the transport of an RF
heated anisotropic temperature species and an isotropic species with
the same effective temperature is expected to scale
similarly. However, at energies only a few times the bulk temperature
(where most of the turbulent transport occurs) the pitch-angle
dependence of the transport is non-trivial: at different energy ranges
it can be weighted towards small or large pitch angles \cite{zhang10}.

For the non-fluctuating distribution function $\fzm(v,\rho)$ in the
gyrokinetic modeling we use a simple isotropic analytical model
derived by Stix, given by Eq.~(33) of Ref.~\cite{stix}.
$\fzm$ depends on radius $\rho$ through the spatial variation of the
absorbed power density, and the densities and temperatures of the
various non-trace particle species.  

By using a non-fluctuating distribution function that is derived by
balancing collisions and quasilinear diffusion due to the interaction
with the RF waves means that we implicitly assume that the radial
transport does not affect the distribution function. In reality the
collisional slowing-down time and the radial turbulent transport time
can compete at certain energies. The study of such non-perturbative
effects is outside the scope of the present paper.

\section{Turbulent transport}
\label{turb}

We calculate the turbulent transport of the heated minorities using
radially local, electrostatic, nonlinear turbulence simulations using
the ``alphas'' branch \cite{alphas} of the gyrokinetic code \gs
\cite{gs2old,gs2new}.  This code is capable of handling a species with
a non-fluctuating distribution which is isotropic but
non-Maxwellian. We will assume the heated species to be present in
trace quantities in the sense that it does not affect the
turbulence. For electrostatic simulations of alpha particles the trace
approximation was shown to be justified in
Refs.~\cite{AngioniPeeters08,wilkie}, although suprathermal pressure
gradients have recently been observed to have an effect in some
electromagnetic simulations \cite{citrin13}.

A trace species is passively advected by the turbulence and it does
not affect the potential fluctuations, thus, given the fluctuation
field, the gyrokinetic equation for such species is linear in the
driving gradients in the distribution function. Consequently the
fluxes in velocity space and configuration space are also linear in
these gradients which may be utilized to calculate the fluxes for an
arbitrary distribution function as shown in Ref.~\cite{wilkiethesis},
and outlined below.  Here we are only concerned with the radial
transport, thus it is instructive to define the energy dependent
radial particle flux as
\begin{equation}
\Gamma(E)=\left\langle \sum_{\sigma} \int \frac{\pi B
  \, d\lambda}{\sqrt{1-\lambda B}}
h_m\langle\vve\rangle_{\mathbf{R}m}\cd\na\rho\right\rangle_{t,\rho},
\label{defGamma}
\end{equation}
where $\vve$ is the fluctuating $E\times B$ velocity,
$h_j=\fj-\fzj-e_j\phi \partial_E \fzj$ is the non-adiabatic perturbed
distribution of the trace species that is characterized by its total
distribution $\fj$ and non-fluctuating distribution $\fzj$ (not
necessarily a Maxwellian), with $\phi$ denoting the fluctuating
electrostatic potential, and $\partial_E$ the partial derivative with
respect to the kinetic energy $E(=m_jv^2/2)$. Furthermore,
$\lambda=\mu/E$ with the magnetic moment $\mu=m_j \vpe^2/(2B)$,
$\vpe^2=v^2-\vpa^2$, and $e_j$ and $m_j$ are the charge and the mass of
species $j$, respectively. The summation is done over the sign of the parallel
velocity, $\sigma$, and $\langle\cd\rangle_{\mathbf{R}j}$ represents a
gyro-average holding the guiding center of the species $\mathbf{R}_j$
fixed, while $\langle\cd\rangle_{t,\rho}$ is an average over the flux
surface and a timescale much longer than the decorrelation time of
turbulent structures. Physically Eq.~(\ref{defGamma}) describes the
net flux of particles across a flux surface at a given energy due to
$E\times B$ drift in the fluctuating electrostatic field. The total
particle and heat fluxes are calculated as $\{\Gamma_j,\,
Q_j\}=\sqrt{2}m_j^{-3/2}\int dE \sqrt{E} \Gamma_j(E) \{1,\, E \}$.  The
non-adiabatic distribution is calculated from the gyrokinetic
equation, which, for species $j$ reads
\begin{eqnarray}
\frac{\p h_j}{\p t}+\left(\vpa
\bv+\vv_d+\langle\vve\rangle_{\mathbf{R}j} \right)\cd\na
h_j-C[h_j]\nonumber\\=-e_j\frac{\p \langle \phi\rangle_{\mathbf{R}j}}{\p t} 
\frac{\p \fzj}{\p E}-\langle\vve\rangle_{\mathbf{R}j}\cd \na \fzj,
\label{gkeq}
\end{eqnarray}  
where $\vv_d$ is the magnetic drift velocity, $C$ is a gyroaveraged
collision operator \cite{abelcoll,barnescoll}. The perturbed potential
is calculated from the quasineutrality condition, but since a trace
species does not contribute to the charge density, $\phi$ is
independent of $\fm$ and thus for such species the problem is linear
in $h_m$, as well as the drives $\p_E \fzm$ and $\nabla \fzm$. 
Consequently  the radial particle flux is also linear in these drive terms
\begin{equation}
\Gamma_m(E)=-D_E \frac{\partial \fzm}{\partial E} -D_\rho \frac{\partial
  \fzm}{\partial \rho},
\label{defDs}
\end{equation}
where $D_E$ and $D_\rho$ are energy dependent transport
coefficients. For a minority species with given charge and mass, these
quantities are determined by the properties of the turbulence, which
in turn only depends on the magnetic geometry and plasma parameter
profiles of the non-trace species. Given two appropriately chosen
distribution functions $\fzm^{(1)}$ and $\fzm^{(2)}$ corresponding to
fluxes $\Gamma_m^{(1)}$ and $\Gamma_m^{(2)}$, the resulting linear
system can be inverted to obtain the transport coefficients. In
particular, these distributions can be Maxwellians with different
radial gradients (chosen so that the linear problem is not singular across
the energy range of interest) \cite{wilkiethesis}; this is the
approach we take. Once the transport coefficients are calculated
$\Gamma_m$ can be calculated from (\ref{defDs}) for any $\fzm(E,\rho)$.

\subsection{Gyrokinetic modeling}
\label{gkmodel}
In the following we detail the gyrokinetic modeling using \gs{} to
obtain $\Gamma_m(E)$. We perform radially local simulations about the
radial location $\rho=0.25$, which is very close to the location where
the power absorption of \het{} is the highest in the cases shown in
Fig.~\ref{deposition}.

The simulations include two bulk ion species, (kinetic) electrons and
two hot Maxwellian species in trace amounts which are of the same ion
type but have different gradients. The latter are used to calculate
the transport coefficients in (\ref{defDs}) as described above. For
the heated minority case the bulk ions are H (70\%) and He (15\%), and
the trace species is \het. For the alpha particle case the bulk
species are D(50\%), T (50\%) and the trace species of ${}^4$He.  The
most important local magnetic geometry and plasma parameters at this
position are the following: The safety factor $q=1.27$, the magnetic
shear $s=0.125$, the aspect ratio $R/a=3.29$, the elongation
$\kappa=1.4$ and $d\kappa/d\rho=0.09$, the triangularity
$\delta=0.075$ and $d\delta/d\rho=0.14$,
$a^{-1}dR/d\rho=-0.08$. Furthermore for all non-trace species
$a/L_{nj}\equiv -d (\ln n_j)/d\rho=0$, for the bulk ions
$a/L_{Ti}\equiv -d (\ln T_i)/d\rho=1.509$, for electrons
$a/L_{Te}\equiv -d (\ln T_e)/d\rho=1.178$, the temperature ratio is
$T_e/T_i=1.02$. In the simulations the normalized pressure $\beta$ is
set to 0. Collisions for bulk ions and electrons were accounted for
using the conservative collision operator describing pitch angle
scattering and energy diffusion \cite{barnescoll}.

The simulations used 28 grid points in extended poloidal angle
covering one $2\pi$ segment along the field line, the number of modes
is $72$ in the binormal ($y$) direction and $48$ in the radial ($x$)
direction with a domain size in both directions being $15$ thermal
Larmor radii of the first ion species (in some cases these numbers
were increased to $96$, $72$ and $20$, respectively). The number of
untrapped pitch-angles moving in one direction along the field is $8$
and the number of energy grid points is $24$. The time step is
$0.1\,a/v_1$ with $v_j=v_1$ for the first ion species.

\subsection{Fluxes against a temperature gradient}
\label{fluxgrad}

First we consider the total particle and energy transport of the
heated \het{} species across its region of strongest power absorption,
$\rho=0.2-0.3$. We assume that in terms of the transport of \het{} the
radial variations in the turbulence are weak compared to the changes
in $\fzm$. This is reasonable when the power absorption varies much
more rapidly than the background profiles (compare
Figs.~\ref{profilefig} and \ref{deposition}; in the studied region is
indicated with the vertical bars). Thus we calculate the transport
coefficients of Eq.~(\ref{defDs}) using a single local gyrokinetic
simulation at $\rho=0.25$ and calculate how the fluxes vary due the
radial variation of $\fzm$. 

\begin{figure}
\centering
\includegraphics[width=0.6\textwidth]{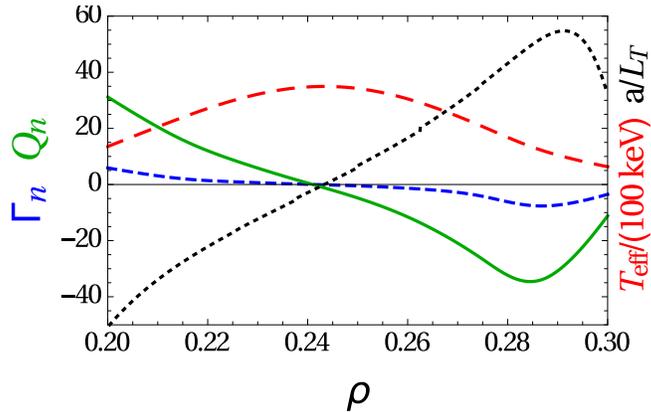}
\caption{Total normalized turbulent particle- (dashed, blue curve) and
  energy (solid, green) fluxes of the heated minority across the
  region of strongest heat absorption.  Absolute fluxes are given by
  $\Gamma_m=\Gamma_n n_m v_r \rho_\ast^2$, $Q_m=Q_n n_m
  v_rT_r\rho_\ast^2$. The effective temperature of the minorities
  ($T_{\rm eff}$ in $\rm 100 keV$; red, long dashed) and the
  corresponding logarithmic temperature gradient ($a/L_T=-d \ln T_{\rm
    eff}/d\rho$; black, dotted) are also plotted. We assumed $P_{\rm
    inj}=15\,\rm MW$ and a power deposition profile shown with the
  solid line in Fig.~\ref{deposition}. Note that positive fluxes are
  flowing radially outward, which means that both of the fluxes are
  flowing against the effective temperature gradient (the radial
  density gradient of the minorities is zero). }
\label{radialtotal}
\end{figure}

Figure~\ref{radialtotal} shows the normalized total particle-
($\Gamma_n$, dashed curve) and energy fluxes ($Q_n$, solid) of \het{}
as a functions of $\rho$, together with the effective temperature of
the heated species ($\teff=n_m^{-1}\int d^3v (m_mv^2/3) \fzm$ given in
$100\,\rm keV$ units, long dashed), and its logarithmic gradient
$a/L_T=-d(\ln \teff)/d\rho$. The absolute fluxes are given by
$\Gamma_m=\Gamma_n n_m v_r \rho_\ast^2$, $Q_m=Q_n n_m
v_rT_r\rho_\ast^2$, where $T_r$, $v_r$ and $\rho_\ast=\rho_r/a$ are
the reference temperature, reference thermal speed and normalized thermal
Larmor radius, which are set to be those quantities for the first ion
species (H). We note that, for the sharp absorption profile we
consider the effective temperature varies on a rather small spatial
scale. At $\rho=0.25$ a $3.5\,\rm MeV$ trapped \het{}
ion has a typical orbit width of $0.08a$ making the local gyrokinetic
treatment questionable for the most energetic ions. However, as we
will show, most of the turbulent transport occurs at a suprathermal
energy range of $\sim 100\,\rm keV$, where the orbit width is
significantly smaller, $\sim 0.01 a$.

Following the radial variation of the absorbed power, the effective
temperature peaks close to the flux surface $\rho=0.25$. Notably, for
our baseline parameters the effective temperature of the heated
species exceeds $3\,\rm MeV$ (while it is expected to be lower for a
less hot plasma). Even more remarkable is the fact that both the
particle and the energy fluxes are flowing in the direction of the
$\teff$ peak (note that radially outward fluxes are defined to be
positive, and the density gradients are zero for all
species). Particularly, the fluxes and $a/L_T$ change sign at the same
radial location.

\begin{figure}
\centering
\includegraphics[width=0.47\textwidth]{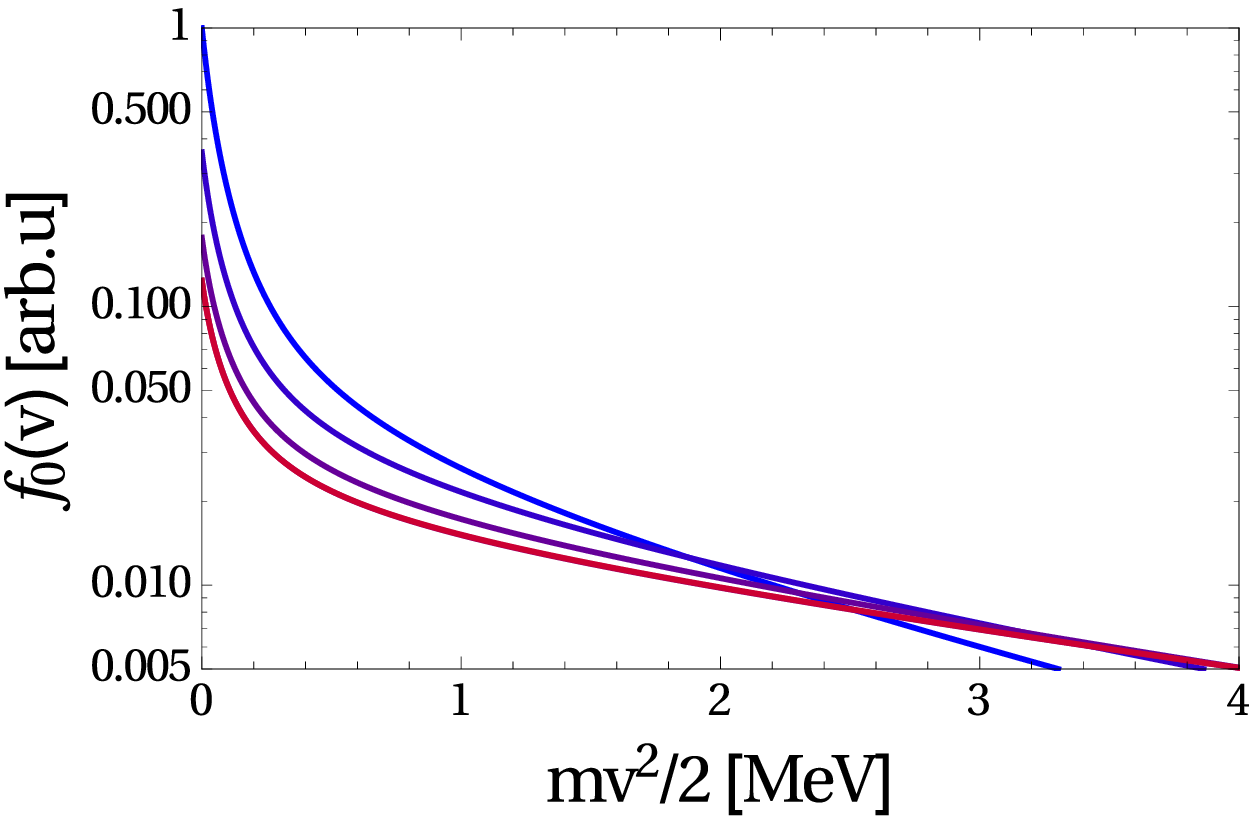}
\put(-35,115){\large (a)}
\put(180,115){\large (b)}
\includegraphics[width=0.47\textwidth]{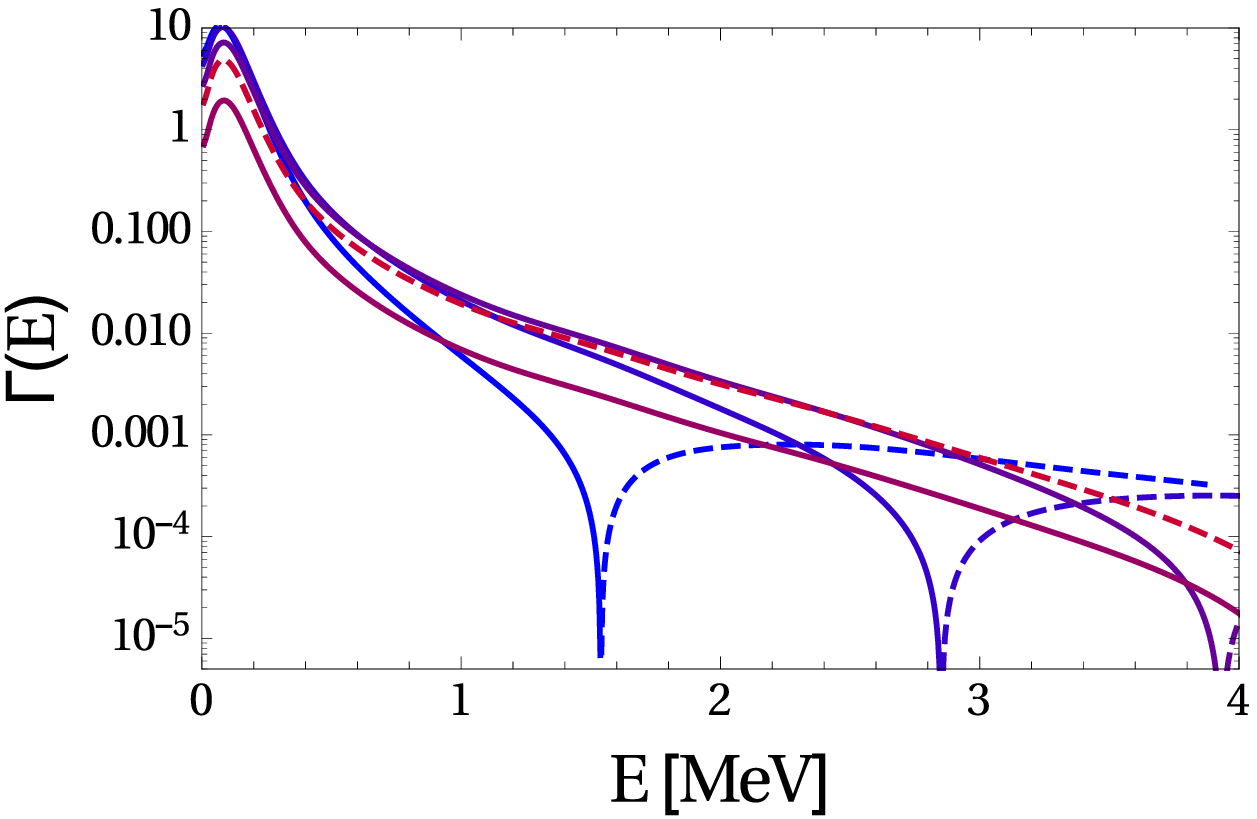}
\caption{Energy variation of $f_0(v)$ (a) and $\Gamma(E)$ (b) at
  different radial locations close to the absorption peak. Dashed
  lines represent negative values. The color code varies from blue to
  purple shades with increasing radius; $\rho$ changes on the scale
  $\{0.2,\,0.2125,\,0.225,\,0.2325,\,0.25\}$. In (a) the $\rho=0.2325$
  and $0.25$ curves, located just below and above the effective
  temperature maximum, overlap almost completely, while the fluxes in
  those locations have opposite sign, as seen in (b). }
\label{theexplanation}
\end{figure} 

The reason for both $\Gamma_m$ and $Q_m$ flowing against the driving
temperature gradient -- which might be puzzling for the first sight --
is illustrated in Fig.~\ref{theexplanation}. The energy dependence of
the distribution function $\fz$ is shown for various radii in
Fig.~\ref{theexplanation}a moving from $\rho=0.2$ to $0.25$,
approaching the $\teff$ peak and passing its maximum slightly at the
last location (henceforth the species subscript for the hot trace
species is suppressed).  The normalization preserves the relative
magnitude of the distributions (i.e. their number density is radially
constant). The effective temperature increases as the peak is
approached radially -- while the \het{} density remains constant --
causing the distribution function to spread out towards larger
energies and become depleted at low energies. The corresponding
$\Gamma(E)$ functions are plotted in Fig.~\ref{theexplanation}b (with
similar color coding for the different radii; dashed lines
representing negative values). The energy dependent fluxes have a
maximum at very low energy ($\sim 100 \,\rm keV$) compared to the
typical effective temperatures ($\sim\rm MeV$) and they decay rapidly
due to FLR effects. The fluxes are thus dominated in an energy range
where the distribution function is depleted when moving radially
towards the higher temperature region. In the light of these
observations fluxes towards the peak of the $\teff$, shown in
Fig.~\ref{radialtotal}, is not surprising: $\fz(E\sim 100 \,{\rm
  keV},\rho)$ has a minimum at the radius of the highest $\teff$.

\begin{figure}
\centering
\includegraphics[width=0.47\textwidth]{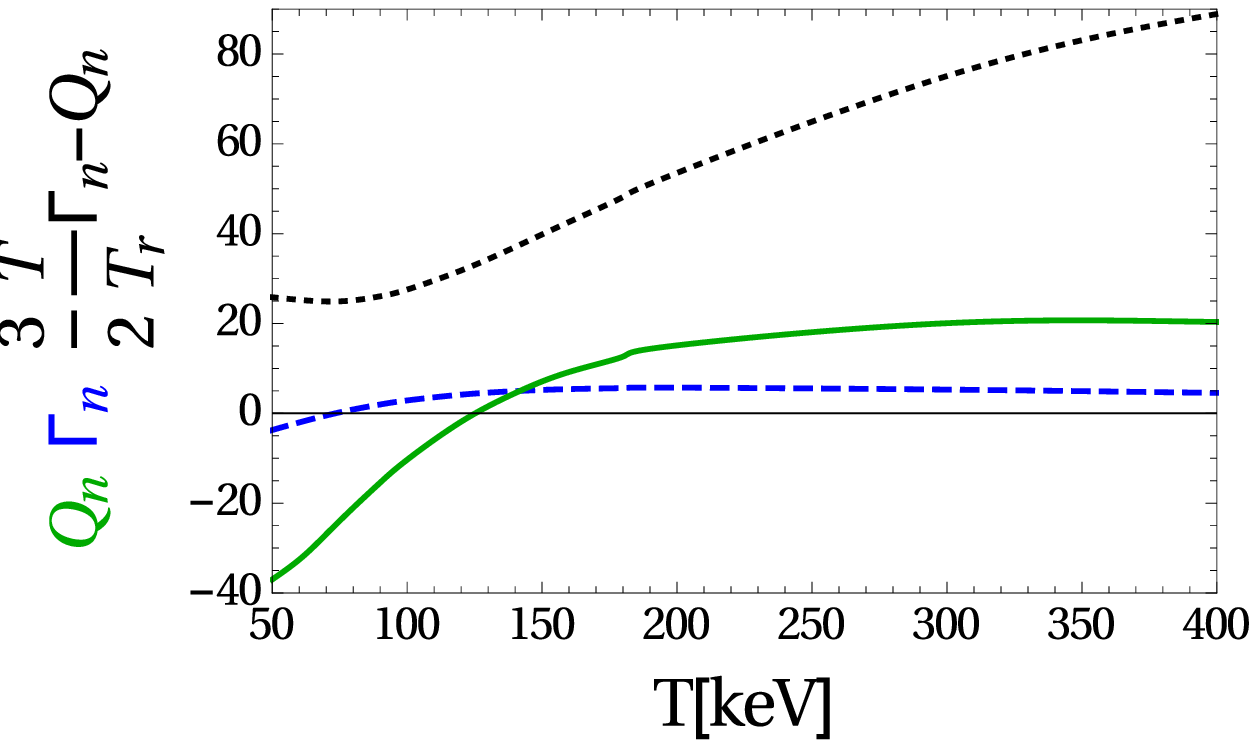}
\put(-35,40){\large (a)} \put(40,40){\large (b)}
\includegraphics[width=0.43\textwidth]{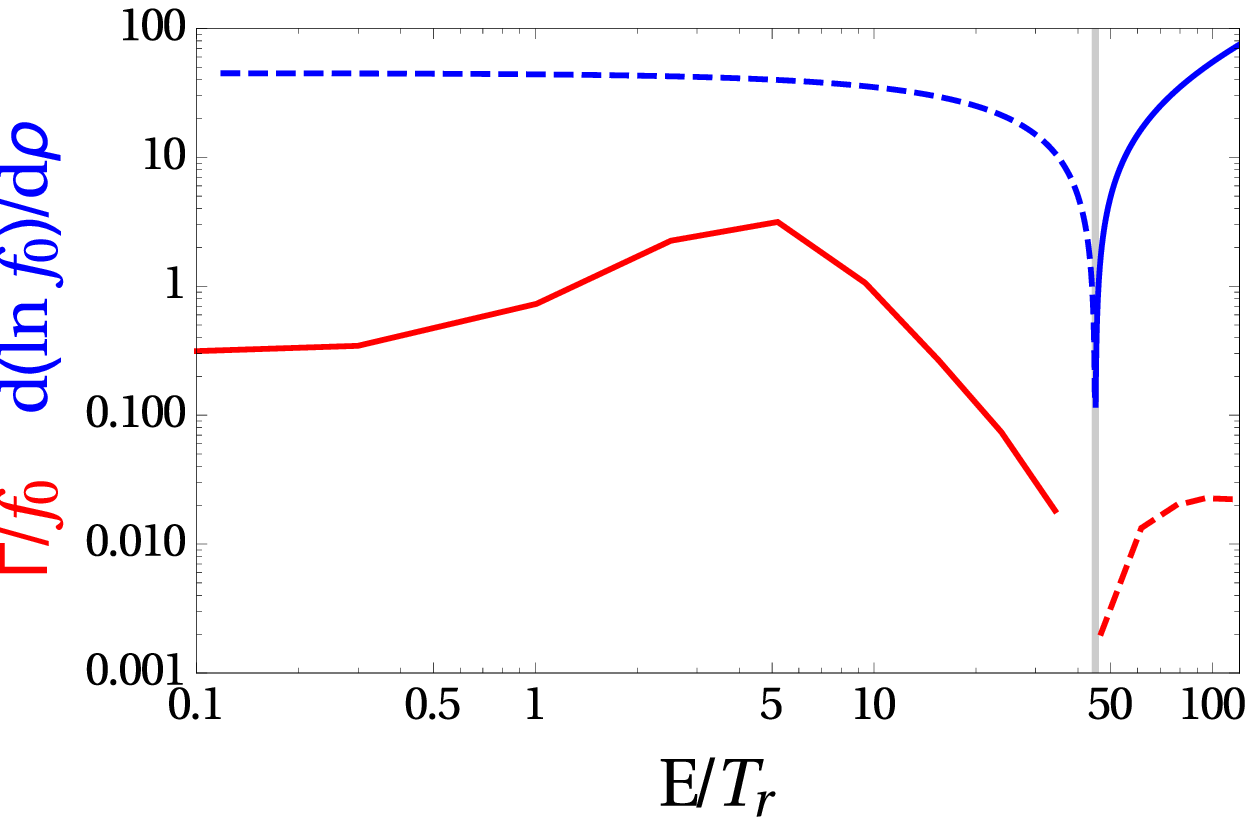}
\caption{Radial fluxes of a high temperature Maxwellian trace. (a)
  Total normalized particle- (blue, dashed curve) and energy (green,
  solid) fluxes as functions of the minority temperature, $\rm T$. A quantity
  related to the entropy production by radial fluxes in this scenario
  (temperature profile peaks outward, density gradient zero) is also
  plotted (black, dotted). Note that while the fluxes change sign with
  increasing minority temperature, the entropy production always stays
  positive. (b) $\Gamma(E)$ (red, lower curve) and $d(\ln f_0)/d\rho$
  (blue, upper curve) as functions of energy for a fixed minority
  temperature ($T/T_r=30$). Both quantities change sign at the same
  energy (dashed curves represents negative values), thus at a given
  energy the flux is not flowing against the radial gradient.}
\label{maxwellianfig}
\end{figure}

It is clear that the phenomena is not the result of the $\fz$ being
non-Maxwellian. Figure~\ref{maxwellianfig}a shows the normalized
particle and heat fluxes (dashed and solid lines, respectively) as
functions of the temperature of a Maxwellian trace species. We assume
that the turbulence properties in terms of $D_E$ and $D_\rho$ are the
same as for our baseline case (detailed in Sec.~\ref{gkmodel}), with
$T_r=25\,\rm keV$, $a/L_{Tm}=-30$, and $a/L_{nm}=0$. For $a/L_{Tm}<0$
(minority temperature increases with radius) we normally expect an
inward heat flux $Q_n<0$; indeed when $T_m\sim T_r$ both $Q_n$ and
$\Gamma_n$ are negative. However, as $T_m$ is increased first the
particle then the heat flux change sign opposing the minority
temperature gradient. Radial fluxes in the presence of gradients
represent sources in the local free energy balance equation (see
\cite{abelGK} and references therein). As a trace species does not
exchange energy with the fields the free energy is conserved for the
species in isolation, and destroyed only by collisions (however small
the collision frequency may be). Therefore negative free energy
generation would imply negative entropy production that is
unphysical. This problem does not arise in our case, since -- for zero
gradients of density and toroidal rotation frequency, and $a/L_{Tm}<0$
-- the free energy generation by the fluxes is proportional to $(3/2)
\Gamma_m T_m -Q_m \propto (3/2)\Gamma_n T_m/T_r -Q_n$. This quantity
remains positive even though the fluxes change signs with increasing
$T_m$ as shown by the dotted curve in
Fig.~\ref{maxwellianfig}a. Considering the energy dependence of the
radial flux and the radial gradient (shown by red and blue curves in
Fig.~\ref{maxwellianfig}b, respectively; negative values are
represented by dashed lines) reveals that the radial flux indeed flows
down the radial gradient at a given energy (the two quantities have
opposite signs and change sign together). In this case the radial
fluxes are dominated by the contribution from an energy range
$E/T_r\approx 5$, where the distribution function is decreasing
radially, while $dT_m/dr$ increases.

\subsection{Transport of heated species and alphas}
\label{transportresults}

Finally, we turn our attention to the radial transport of the hot
minorities and compare the results for the heated impurities and alpha
particles. We consider the radial location $\rho=0.275$, somewhat
outside the maximum of the RF power deposition so that the effective
temperature of~\het{} decreases radially while it is still very high
($T_{\rm eff}\approx 2.8 \,\rm MeV$).

\begin{figure}
\centering
\includegraphics[width=0.46\textwidth]{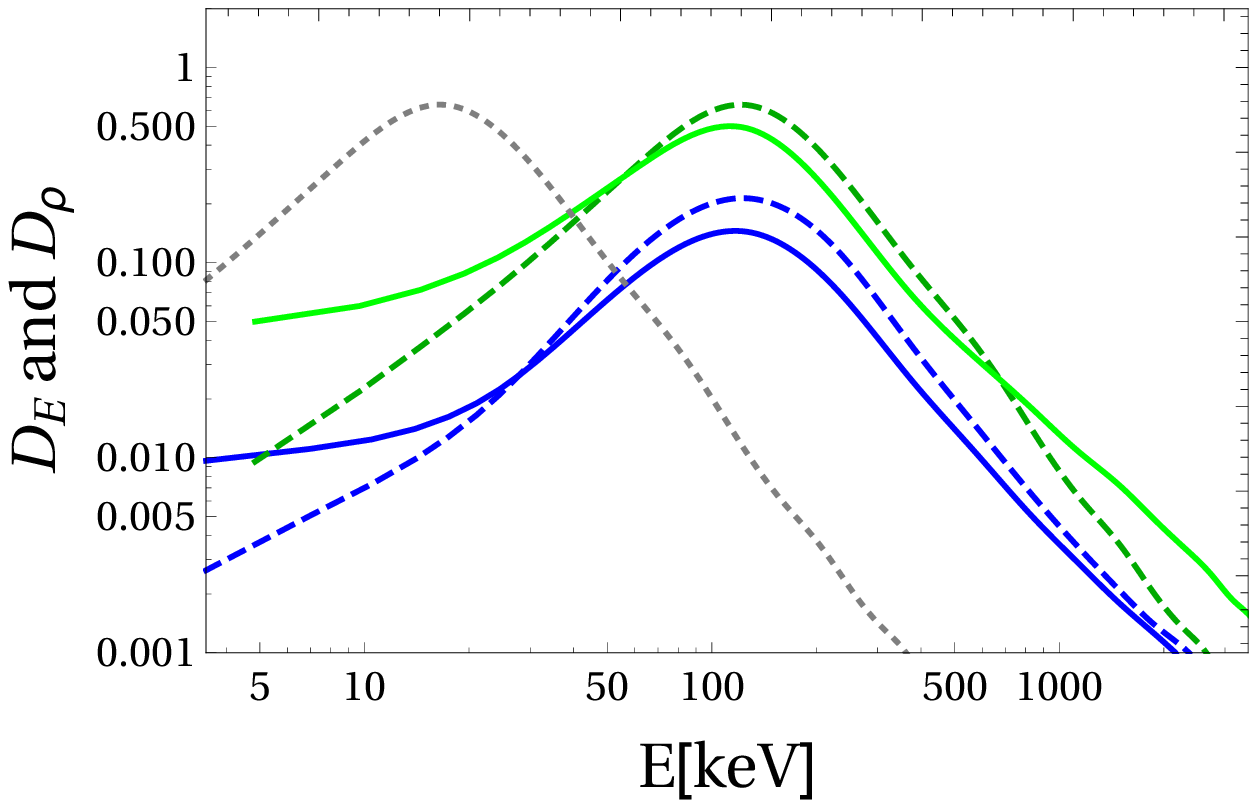}
\put(-30,110){(a)}
\put(172,110){(b)}
\put(-145,92){\normalsize $D_\rho$}
\put(-167,70){\normalsize $D_E$}
\includegraphics[width=0.45\textwidth]{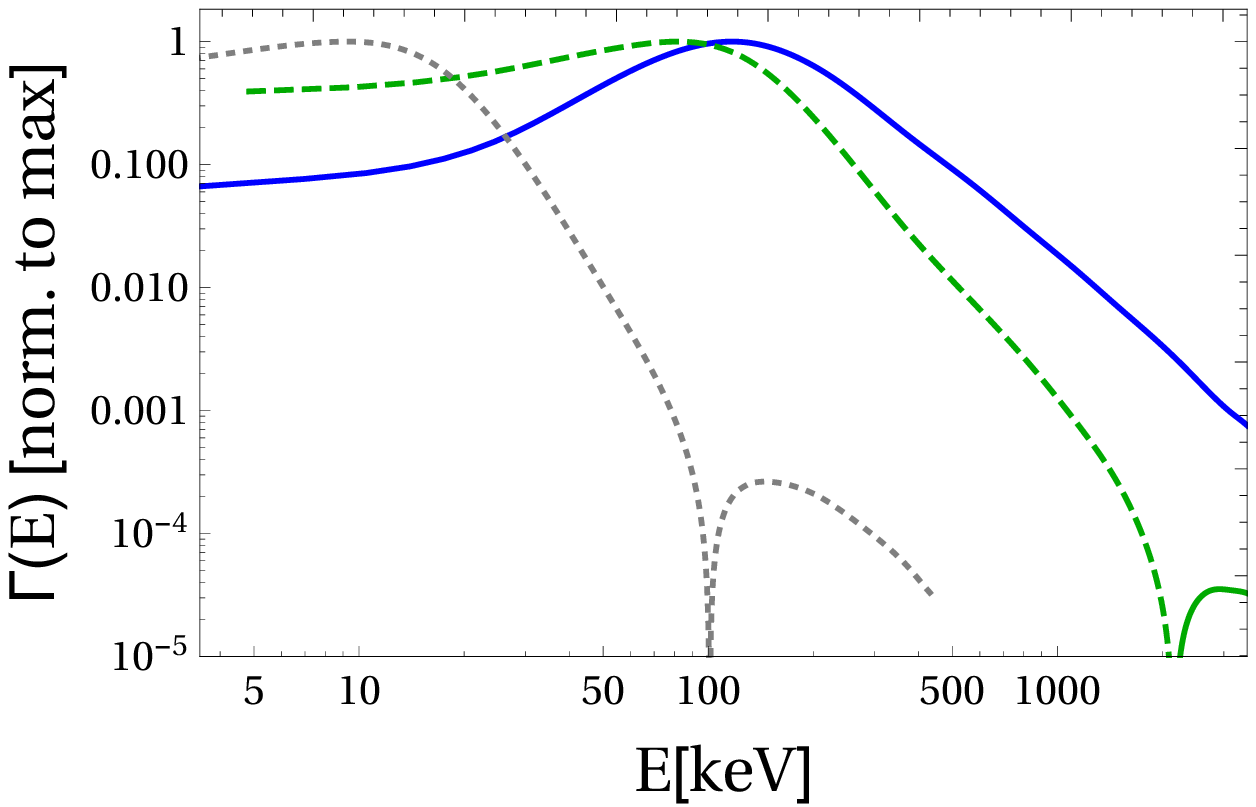}
\includegraphics[width=0.46\textwidth]{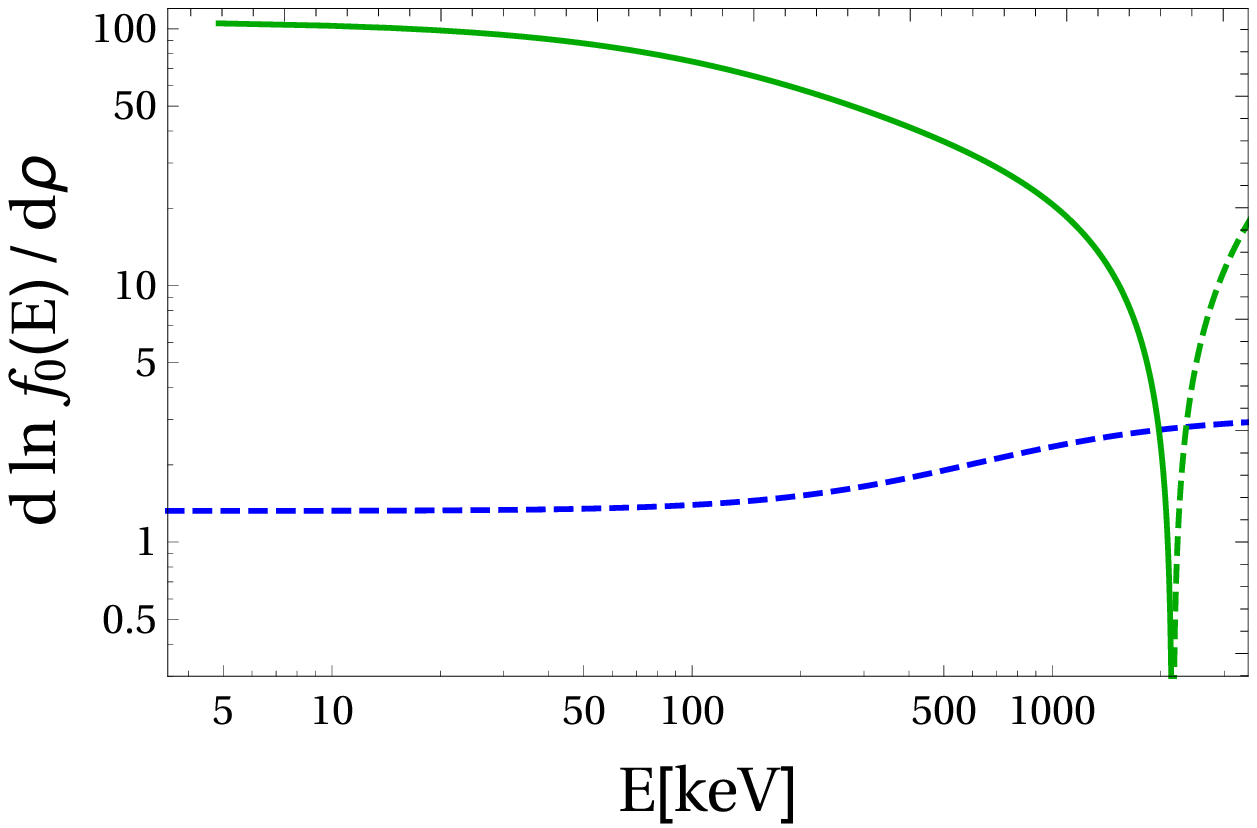}
\put(-30,120){(c)}
\put(172,120){(d)}
\includegraphics[width=0.45\textwidth]{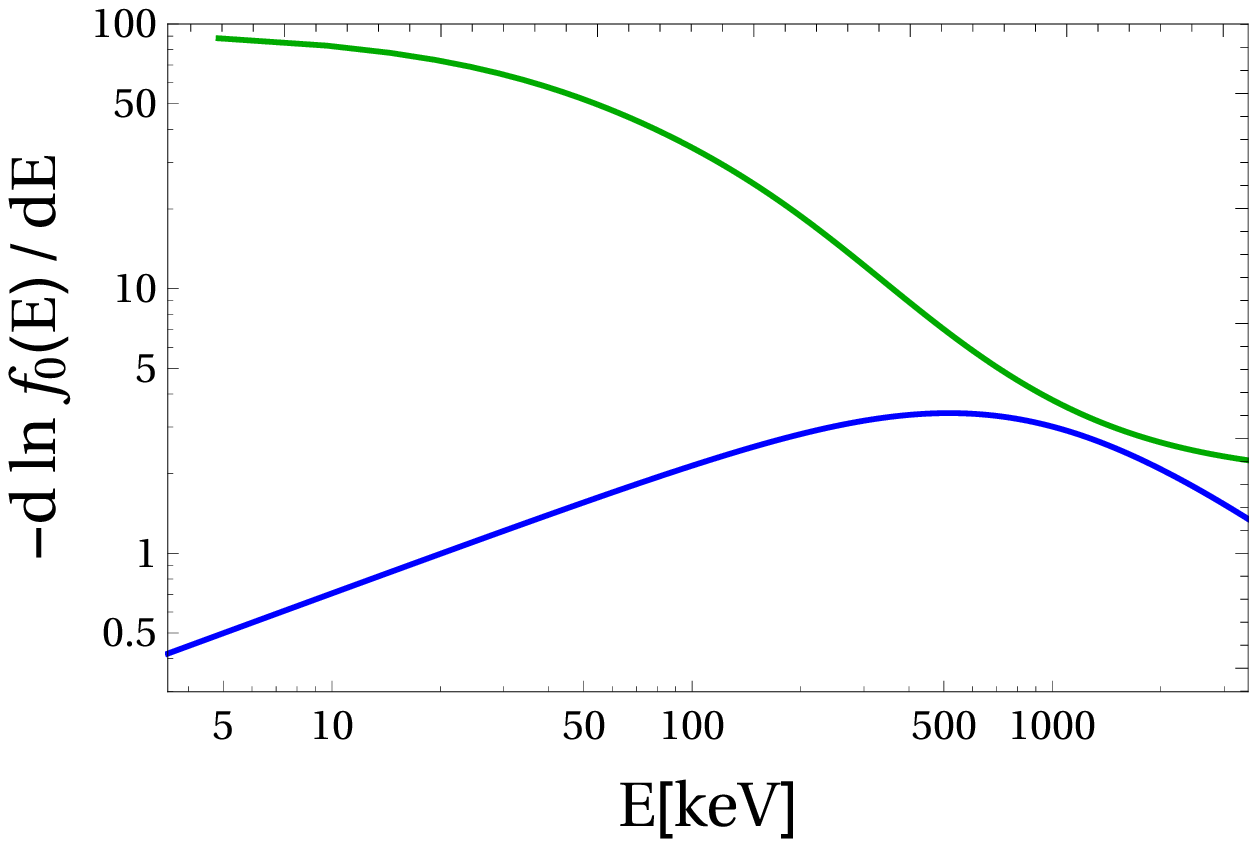}
\caption{Energy dependent transport coefficients (a), radial particle
  flux (b), and logarithmic energy- (c) and radial (d) derivatives of
  the distribution functions of the energetic minority. The
  calculations are based on the 20020100 ITER discharge at
  $\rho=0.275$. Blue curves: $\alpha$ slowing-down distribution,
  $\alpha$-source and turbulence calculated for $50-50\%$ D-T. Green
  curves: heated $\rm He^3$, ion composition and heating of
  Fig.~\ref{deposition} (solid line). Dashed lines represent negative
  values. Dotted lines show results for the reduced density and
  temperature plasma corresponding to the dashed curve of
  Fig.~\ref{deposition}; these are $|D_E|$ in (a) and $|\Gamma|$ in
  (b).  }
\label{comparisonfig}
\end{figure}

Assuming similar background profiles (discussed in
Sec.~\ref{heatedspecies}) but different ion composition results in
qualitatively similar turbulent transport in terms of the transport
coefficients for a trace species. The transport coefficients
$D_E$ (dashed curves, given in units of $\rho_\ast^2v_rT_r$) and
$D_\rho$ (solid curves, in units of $\rho_\ast^2v_r$) are shown
for alpha particles (blue curves, D-T plasma) and heated \het{} (green
curves, H-He plasma) in Fig.~\ref{comparisonfig}a.  In terms of
their energy dependence, both quantities peak around $100\,\rm
keV$ and decay towards high energies due to FLR effects. The radial
coefficient $D_\rho$ is positive, as it should be for positive entropy
generation.  The energy coefficient $D_E$ is negative, similarly to
the findings of Ref.~\cite{waltz13}. It being negative means that
without radial variations in the distribution function there would be
inward fluxes (except for distributions non-monotonic in
  energy), which can be thought as a generalized pinch due to the
background turbulence, driven by gradients in the distribution of the
bulk species. However, $D_E \partial_E f_0\ll D_\rho
  \partial_\rho f_0$, that is, in terms of $\Gamma(E)$ the radial
  variation of the distribution function at a given energy is more
  important than its energy variation. This is expected for a hot
  species with radial gradients comparable to, or larger than, those
  of the bulk \cite{chen16}: Although the diamagnetic frequency of the
  bulk species is comparable to the characteristic mode frequency of
  the underlying drift waves, the diamagnetic frequency of the hot
  species -- ultimately giving rise to the $D_\rho \partial_\rho f_0$
  contribution -- is typically much higher than the mode frequency.
  The transport coefficients of the heated ${}^3$He are approximately
  $3$ times larger in absolute magnitude than those of the alpha
  particles, due to differences in the bulk ion composition. However,
  this difference merely reflects a change in the turbulence
  intensity: a similar difference in the total ion heat fluxes is also
  observed. These simulations are gradient driven and exhibit a stiff
  ion heat transport; in an experiment the gradients would slightly
  adjust to produce a fluctuation amplitude and a heat flux as
  required by the sources. Thus, for a similar heating power inside
  the flux surface of interest we expect the magnitude of the
  transport coefficients to be very similar in the two plasmas. 

As we have established that turbulence has a similar effect on the hot
species in both plasmas, any notable differences in terms of their
radial transport should come from differences in the their
distribution functions. The generation mechanism of the hot species is
fundamentally different: while alpha particles are born at high energy
with a radially varying source and slow down due to collisions, the
distribution of the heated species is shaped by a quasilinear
diffusion due to the interaction with the electromagnetic field and
collisions. In particular, if we could instantaneously ``switch on''
D-T fusion, alpha particles would start filling up phase space from
high energy, while when RF heating is switched on the distribution
expands from lower towards higher energies. Naturally, the hot alpha
particle distributions have a radial density variation due to the
source gradients, while the heated distribution can have a potentially
strong effective temperature gradient depending on the power
deposition profile.

With time the fluxes in phase-space would become divergence free as
the minorities would settle at some equilibrium distribution function
(analogous to the converged plasma parameter profiles in predictive
modeling). In our gradient driven modeling we cannot address this
question, and we are only concerned with the fluxes for given
distribution functions. The radially decreasing source strength of
alphas and the radially decreasing effective temperature thus
clarifies $d (\ln f_0)/d \rho$ being negative for alphas for all
energies, and being large and positive for \het{} over most of the
energy range considered, as shown in Fig.~\ref{comparisonfig}c. 

The energy derivatives are shown in Fig.~\ref{comparisonfig}d. The
high and low energy limits of the \het{} distribution are different
temperature Maxwellians thus $-d(\ln f_0)/dE$ varies between two
constant asymptotes, while this quantity for the alpha slowing down
distribution is non-monotonic with a peak comparable to the critical
energy (the He ash distribution is not considered). The resulting
energy dependent fluxes normalized to their maximum values are shown
in Fig.~\ref{comparisonfig}b. For the heated \het{} both the energy
and the radial gradient drive terms cause inward fluxes, thus the
total fluxes are inward. A sign change of $\Gamma(E)$ occurs near
where $d \ln f_0(E)/d\rho$ changes sign, but above that energy
$\Gamma(E)$ is so small that the total particle and energy fluxes
remain negative.  This would, on longer time scales, lead to an
accumulation of the heated species around the maximum of the power
absorption. While the energy derivative contribution is negative for
the alphas too, the radial gradient driven part is positive for them,
producing outward total flux.

After considering a completely flat minority density profile and a
rather sharp effective temperature variation in our heated minority
example, now we consider the effect of these assumptions being
relaxed. At the radial location $\rho=0.275$ the effective temperature
gradient is $a/L_T=35.6$, which is due to a comparably large gradient
in the absorption profile $a/L_P\equiv -d(\ln P_{\rm
  abs})/d\rho=23$. Reducing $a/L_T$ at zero density gradient linearly
scales down the fluxes, and their sign is unaffected. However, when
the minority has a density gradient comparable to the effective
temperature gradient, the fluxes can change sign. In our heated
minority example, if $T_{\rm eff}$ and $a/L_T$ are held fixed
$\Gamma_m$ ($Q_m$) is found to change sign at a logarithmic minority
density gradient of $a/L_{n}=48$ ($a/L_{n}=28$). If the absorption
profile is held fixed then the sign change happens at lower density
gradients, $a/L_{n}=18$ ($a/L_{n}=13.5$), since then the density variation
also affects the effective temperature variation.

Although we do not evolve the distributions of the hot trace species
towards a steady state, we note that such task is feasible for alpha
particles, and has been done with the {\sc t3core} code presented in
Ref.~\cite{wilkiethesis,tthreecore}. For alphas the steady state is
determined by the sources, the slowing down and the radial transport
due to turbulence; the generation through fusion should be balanced by
the total radial transport. The heated species problem is considerably
more involved, since the process of the heating itself depends on the
heated distribution function, we do not attempt to tackle this very
challenging problem here.

For the \het{} heating case we have shown simulation results with
plasma parameters similar to a projected ITER DT discharge (Case
1). Although the similarity of the plasma parameter profiles was
convenient in simplifying the comparison, the densities and
temperatures in a non-activated phase discharge relevant for a TIM
heating experiment could be significantly lower. To asses
corresponding differences in the transport we calculate the transport
coefficients and the radial flux of the heated species for densities and
temperatures scaled by constant factors (see caption for
Fig.~\ref{deposition}, Case 2). Also, we reduce the injected power to
$P_{\rm inj}=10\,\rm MW$, since the power absorption by the \het{} ions is
more effective at the reduced bulk temperatures.

The local gyrokinetic simulation results are only affected by these
parameter changes through changes in collisionality (besides changes
in the normalization, which do not enter the simulation). The
collisionality in this ion temperature gradient mode driven turbulence
has only minor effects. The transport coefficients are thus similar,
apart from a shift in the energy range by a factor $7.3$ that is the
same as the ratio between the temperatures in Case 1 and Case 2. For
comparison we show $|D_E|$ for Case 2 in \ref{comparisonfig}a, which
is to be compared with the green dashed line for Case 1. Due to the
change in collisionality the heated distribution in Case 2 is not
simply down-shifted in energy but it is slightly different, which
leads to that $\Gamma(E)$ is visibly different between the two cases
(see $|\Gamma(E)|$ plotted with dotted line in \ref{comparisonfig}b),
with the ratio between the peak energies being $11.2$ and the ratio
between the energies where the sign change occurs is $21.4$. The
qualitative behavior and the sign of the total fluxes remain the same.


\section{Conclusions}
\label{conclusions} 

We have studied the turbulent transport properties of various
energetic, non-Maxwellian trace species in magnetized plasmas. In
particular we compared the radial transport of the hot species in the
three ion minority ICRH scheme (assuming a peaked off-axis power
deposition) to that of alpha particles based on projected ITER plasma
parameter profiles. The motivation for such comparison is the prospect
of mimicking alpha particle confinement with $\rm MeV$ range ions
already in the non-activated phase of ITER operation, or in
  present day experiments.

As a general observation, we find that a species (Maxwellian or not),
characterized by an effective temperature much higher than the bulk
temperature and an effective temperature gradient sufficiently higher
than the density gradient, can develop energy and particle
fluxes flowing against the gradients. The reason is that most of the
turbulent transport occurs on an energy range comparable with the bulk
temperature, and it is FLR-suppressed towards high energies. A
temperature gradient for such a high temperature species corresponds
to a radial gradient of the distribution function with the opposite
sign at these energies much lower than the species' effective
temperature. The behavior of heated minorities is governed by
  this phenomena. For minorities much hotter than the bulk plasma the
use of energy dependent radial particle flux is more informative than
the total particle and heat fluxes.

We find that the energy dependent turbulent transport
coefficients of the passively advected species are similar in the
different ion configurations considered (H-He plasma with a trace
\het{} and D-T plasma with trace alphas). The radial turbulent
transport is dominated by a contribution from a suprathermal energy
range, $\sim 100 \,\rm keV$ for the ITER-relevant bulk temperatures
considered. However, the radial particle transport is different
because of differences between the distribution functions of the hot
species. The alphas have a radially varying particle source
corresponding to a radially decreasing density, which necessarily
corresponds to an outward total particle transport in steady sate. The
heated minorities, if they originally have a density profile less
steep than the effective temperature profile, will be radially
transported towards the region where most of the RF heat deposition
takes place, as their distribution function is depleted there across
the energy ranges dominating the turbulent transport. These results
suggest that their steady state density profile should be peaked in
the region where the heat deposition is the strongest. Despite the
observed differences in their turbulent transport, by generating ions
in the right energy range in a controllable fashion, the TIM scheme
would still be very useful in validating alpha particle transport
prediction tools before the activated phase of the ITER operation. Our
results point out the importance of these tools to account for effects
stemming from the differences in alpha and heated minority
distribution functions.


\ack The authors are grateful to P.~Helander, A. Schekochihin and
I.~Abel for fruitful discussions, and to M.~Brambilla and R.~Bilato
for providing the TORIC code. This work was supported by the
International Career Grant (Dnr.~330-2014-6313) and the Framework
grant for Strategic Energy Research (Dnr.~2014-5392) from
Vetenskapsr{\aa}det, and the US DoE grants DEFG0293ER54197 and
DEFC0208ER54964. The gyrokinetic simulations were performed on Hopper
at NERSC, and on Hebbe at C3SE (project nr. SNIC2016-1-161).


\section*{References}
\bibliographystyle{iopart-num}
\bibliography{HeatedMinority1}



\end{document}